\documentclass[letterpaper,twocolumn,10pt]{article}
\usepackage{usenix2022_SOUPS}

\usepackage{tikz}
\usepackage{amsmath}

\usepackage{filecontents}

\usepackage{color}
\usepackage{siunitx}
\usepackage{booktabs}
\usepackage[normalem]{ulem}
\usepackage{textcomp}
\usepackage{hyperref} 
\usepackage{fontawesome}
\usepackage{url}
\usepackage{colortbl}
\usepackage{graphicx}
\usepackage{subfigure}
\usepackage{array}
\usepackage{amssymb}
\usepackage{pifont}
\newcommand{\cmark}{\ding{51}}%
\newcommand{\xmark}{\ding{55}}%
\usepackage{tikz}
\usepackage{CJK}
\usepackage[normalem]{ulem}
\usepackage{verbatim}

\usepackage{tcolorbox}
\usepackage{CJK}
\usepackage[normalem]{ulem}
\definecolor{mygray}{gray}{.95}
\definecolor{myblue}{RGB}{18,81,149}
\newcounter{consideration}

\newcommand{\rev}[1]{{\color{black}#1}}

\begin{document}

\date{}

\title{\Large \bf Iterative Design of An Accessible Crypto Wallet for Blind Users}




\author{
{\rm Zhixuan Zhou$^{\ast}$}\\
University of Illinois at Urbana-Champaign
\and
{\rm Tanusree Sharma$^{\ast}$}\\
University of Illinois at Urbana-Champaign
\and
{\rm Luke Emano}\\
University of Illinois at Urbana-Champaign
\and
{\rm Sauvik Das}\\
Carnegie Mellon University
\and
{\rm Yang Wang}\\
University of Illinois at Urbana-Champaign
}

\maketitle
\def\thefootnote{*}\footnotetext{The first two authors contributed equally to this work.}\def\thefootnote{\arabic{footnote}}


\begin{abstract}
    Crypto wallets are a key touch-point for cryptocurrency use. People use crypto wallets to make transactions, manage crypto assets, and interact with decentralized apps (dApps). However, as is often the case with emergent technologies, little attention has been paid to understanding and improving accessibility barriers in crypto wallet software. 
    We present a series of user studies that explored how both blind and sighted individuals use MetaMask, one of the most popular \rev{non-custodial} 
    crypto wallets. We uncovered inter-related accessibility, learnability, and security issues with MetaMask. We also report on an iterative redesign of MetaMask to make it more accessible for blind users. This process involved multiple evaluations with 44 novice crypto wallet users, including 20 sighted users, 23 blind users, and one user with low vision. Our study results show notable improvements for accessibility after two rounds of design iterations. Based on the results, we discuss design implications for creating more accessible and secure crypto wallets for blind users. 
\end{abstract}

\vspace{-2mm}
\section{Introduction}
\vspace{-2mm}

\label{sec:intro}

Crypto wallets are an essential touch point for users to interact with blockchain and cryptocurrency technologies.
These wallets are user interface wrappers over public/private key pairs that allow users to securely store, send, receive, and monitor their digital assets without mastering the underlying blockchain technology or running their own blockchain nodes~\cite{wallet}.
In addition, they facilitate authenticating into and interacting with decentralized applications (dApps)~\cite{antonopoulos2018mastering, dapp}, 
and simplify the participation in the governance of Decentralized Autonomous Organizations (DAOs)~\cite{thyagarajan2021reparo}.
In short, end-user use of cryptocurrency and blockchain technologies is synonymous with the use of crypto wallets.



Unsurprisingly, understanding and improving the end-user experience with crypto wallets has been the subject of much prior research \cite{albayati2021study,frohlich2021don, ghesmati2022usability, voskobojnikov2021u}.
Yet, as is common with emergent and rapidly evolving technologies \cite{das2022security}, ensuring the accessibility of crypto wallets has not been a focus of this prior effort.
This lack of accessibility, in turn, effectively marginalizes blind users from participating in the emerging ecosystem of cryptocurrency and blockchain apps~\cite{marginalize}.
In response, Marta Piekarska, Director of Ecosystem at Hyperledger, highlighted the importance of accessible wallets for the blind community and how current wallets have not addressed their needs \cite{unsolved}. The National Federation of the Blind has, likewise, called for improving the accessibility of cryptocurrency technology~\cite{nfb}, noting a lack of centralized oversight could cause poor accessibility outcomes. 


Moreover, accessibility is closely tied to usability and security.
Usability issues often disproportionately impact blind users~\cite{usability}.
For example, while all novices may be overwhelmed by the highly technical concepts foregrounded by crypto wallets (e.g., seed phrases, transaction gas fees)~\cite{concept}, novices who are blind face the additional burden of confronting these concepts with user interfaces that are inaccessible to screen readers.
Likewise, security concerns are rampant in the Web3/crypto space, particularly phishing scams that trick users into sharing their private key or seed phrase with attackers~\cite{phishing} and sending their cryptocurrency assets to an attacker's address~\cite{screen}. 
Blind users may be even more at risk than others: prior work has shown that accessibility tools can help blind users detect and protect themselves against phishing attacks~\cite{blythe2011f}, a tactic that cannot be executed when crypto wallets themselves are inaccessible.

We present a multi-phased, iterative re-design of a popular crypto wallet, MetaMask, to both examine and improve accessibility of  crypto wallets. \rev{MetaMask is a non-custodial wallet, meaning that end users are responsible for managing their own private keys. We focused on MetaMask because it was the dominant wallet when we conducted this work in 2022, with 30 million monthly active users \cite{30M}.} We designed the resulting wallet, iWallet, with ``inclusiveness'' in mind, aiming to improve accessibility for blind users.   
Specifically, our work was guided by the following research questions: 

\noindent \textbf{RQ1}: What are the experiences of blind users with current crypto wallets? \\
    \textbf{RQ2}: How can crypto wallets be made more accessible to blind users?

To answer RQ1, we collected and analyzed user reviews of 10 existing crypto wallets, performed a competitive analysis of the accessibility and other usability aspects of them, and conducted a usability study (N=18) of MetaMask (Version 10.23.2). \rev{Accessibility issues were common in all of the 10 popular wallets we analyzed. Eight of the ten wallets in our competitive analysis did not implement any accessibility features. The two that did (MetaMask, Coinbase) still had accessibility issues such as a confusing heading hierarchy, poor contrast between text and background colors, and a lack of keyboard navigation accessibility.}
In the usability study, we observed the behavior of 10 sighted users, 7 blind users, and one user with low vision while they performed basic tasks (e.g., creating an account, making transactions, importing an account) using MetaMask. We aimed to identify accessibility concerns and tasks that were disproportionately difficult for blind users to inform our later accessibility-centered redesign. We uncovered several accessibility issues, including unlabeled and poorly labeled buttons and a lack of confirmation notifications. These accessibility issues also exacerbated a number of correlated usability and security issues. For instance, when creating or importing a wallet account in MetaMask, users had to manually write down and type in their seed phrase, consisting of 12 automatically generated words --- a challenge that was much more difficult for blind participants. 

To answer RQ2, we followed an iterative design process~\cite{nielsen1993iterative} to implement and evaluate a more accessible version of MetaMask: iWallet. 
Based on our findings for RQ1, we designed iWallet with a focus on accessibility --- touching on education, security, and usability. To evaluate our first redesign, we conducted a pilot study with a new set of 10 sighted and 8 blind participants. Building on their feedback, we iterated on our design and conducted a summative evaluation with a new set of 8 blind users. In our final design, we updated several features: we improved button labels, provided text summaries to improve the accessibility of video instructions, and prioritized downloading the seed phrase as a back-up option over manually writing it down.
The summative evaluation confirmed that participants found iWallet more usable and accessible than the original MetaMask.
\rev{On the System Usability Scale (SUS), participants rated iWallet much higher than MetaMask (81 vs 70, out of 100). Much of this improvement could be attributed to reducing complexity, improving ease of use, and reducing the need for prerequisite knowledge in their interactions with the wallet}. 
Our blind participants, in particular, expressed positive feedback overall.
Specifically, they appreciated: (i) the adequate labeling of buttons and web elements as well as the accessible secret recovery phrase\footnote{The secret recovery phrase in iWallet and MetaMask is equivalent to the seed phrase in other wallets. We use these two terms interchangeably.} management process; (ii) being prompted to re-type and confirm the receiving address when sending cryptocurrencies to other accounts to ensure transaction security, since differentiating wallet addresses is harder for blind users; and, (iii) the accessible, video- or text-based explanations of technical crypto concepts (e.g., wallet address).

\rev{Our work makes two main contributions. First, through a competitive analysis of popular crypto wallets and a multi-phase user study, our work is the first to examine and improve the accessibility of crypto wallets for blind users, cataloging accessibility issues faced by blind users when using popular crypto wallets such as MetaMask. One concerning finding is that popular wallets such as MetaMask failed common  accessibility standards such as WCAG.
Second, through an iterative design process, we fix accessibility issues (e.g., by adding labels), introduce new accessibility designs such as downloadable, encrypted seed phrases, and provide key insights for researchers and practitioners on how to design more accessible crypto wallets for blind users.}

\vspace{-2mm}
\section{Related Work}
\vspace{-2mm}

\label{sec:lit}

\vspace{-2mm}
\subsection{UX of Blockchain-Based Apps}
\vspace{-2mm}

 
User experience (UX) is a major challenge in blockchain-based applications, especially for non-technical users who find it daunting to understand the technical aspects of blockchain~\cite{user:experience:ahfe}. Additionally, the security of blockchain technology can be compromised if the UX is not designed with security in mind, leading to security breaches~\cite{user:experience:security}.

Decentralized Autonomous Organization (DAO), a disruptive advancement in blockchain-based applications, achieves algorithmic governance through smart contracts while heavily relying on human collaboration in decision-making~\cite{rikken2022creating, meijer2018governance,marquez2021attempt}. DAOs face usability issues due to the complex nature of the underlying technology, particularly with respect to smart contracts. Users often encounter difficulties managing their tokens to participate in voting and proposals, which is considered challenging for less tech-savvy individuals who may not fully understand the process~\cite{jafar2021blockchain}. Non-fungible token (a.k.a. NFT), another popular blockchain application, also suffers from poor user experience, especially during the onboarding process, which can lead to loss of money and scams~\cite{sharma2022s, wang2021non}. Uniswap is a decentralized cryptocurrency exchange with two main features: cryptocurrency swapping and pooling cryptocurrency as liquidity. A research report indicated that the analytics features were not easily accessible to users navigating the Uniswap app, which could cause confusion and hinder user adoption~\cite{uniswap}. Furthermore, users' mental models, influenced by traditional financial applications such as stock exchanges, could impact their perceived usability and user experience of this decentralized platform. In the case of blockchain-based gaming, players may find the technology not intuitive, with accessibility being a significant barrier~\cite{gaming}. 

\vspace{-2mm}
\subsection{UX of Financial Apps}
\vspace{-2mm}
Since crypto wallets could be considered a type of financial app, we also looked into the prior literature in financial apps, identifying a number of issues related to user experience, including trust, security, usability, and design~\cite{albayati2020accepting,hamilton2005challenges, kiljan2016survey}. Akturan et al. 
conducted a user experience inspection of financial applications, focusing on mobile banking, online trading, and personal financial management ~\cite{akturan2012mobile}. Medhi et al.~\cite{medhi2009mobile} examined mobile banking user interfaces in developing countries and proposed a framework that emphasized the importance of designing for users with limited literacy and numeracy skills to improve accessibility and financial inclusion. Teresa et al.~\cite{teresaglobalisation} 
examined different methods and tools for evaluating the user experience of financial services and highlighted the challenges of conducting research in the highly regulated financial industry for many researchers. Several studies have investigated the relationship between trust and user experience in mobile banking using a combination of surveys and interviews. For example, it was found that trust was a key factor in user experience and that factors such as security, reliability, and transparency could have a significant impact on users' trust in mobile banking applications~\cite{zhou2012examining}. In addition, Wentz et al. and Goundar et al.~\cite{wentz2017exploring, goundar2023exploring}
have highlighted the lack of consideration for accessibility in designing various digital financial services.
\vspace{-2mm}
\subsection{Usability of Crypto Wallets}
\vspace{-2mm}
Cryptocurrency is becoming increasingly popular in recent years. According to an NBC News poll, one in five Americans has invested in, traded, or otherwise used cryptocurrency \cite{one:in:five}.
People hold cryptocurrency for investment purposes, to purchase goods including everyday items, and to learn more about crypto assets out of curiosity. Crypto wallets are software wrappers on top of private/public key pairs to facilitate easy interaction with the underlying blockchain~\cite{key}. 
However, usability issues prevent them 
from reaching the mass public \cite{cryptocurrency, exchange}. Researchers in the HCI community have tried to identify usability issues in crypto wallets \cite{blockchain:hci}. For example, Moniruzzaman et al. adopted an analytical cognitive walk-through inspection with 5 participants, and found many crypto wallets lacked good usability in performing fundamental tasks \cite{usability}.  
From an end user's perspective, a blockchain usability report identified common usability issues of crypto wallets through a survey of over 200 crypto holders \cite{usability_report}. The report found that users had difficulty interacting with wallets and, in turn, the underlying blockchain. 
More than half of the users had at least one concern or problem with their transactions. Many users did not have full confidence in transactions, fearing that something might go wrong. The most reported issue was that users were not sure if a provided wallet address was accurate. Transaction fees were conceptually confusing to the users since the connection between fees and delivery times was often unclear.
Similarly, 6,859 reviews regarding user experience of five mobile crypto wallets were identified and qualitatively analyzed \rev{by Voskobojnikov et al. \cite{usability-chi}.} 
Lack of guidance during the setup made it challenging to create a wallet. 
Qualitative quotes by interview participants were presented \rev{by Voskobojnikov et al.\cite{cryptocurrency}}, pointing out usability issues of popular crypto wallets such as MetaMask: \textit{``You have to enter a gas amount in some other currency that you have never heard called Gwei and then a lot of the times the recommended amount isn't enough.''}

Numerous attacks have been conducted against the blockchain ecosystem \cite{security}, especially decentralized finance (DeFi), such as flash loans~\cite{flash_loan}. 
Such security vulnerabilities could often be attributed to usability issues. Mai et al.~\cite{mental_model} found that users' misconceptions of cryptocurrency and the blockchain were associated with their inappropriate security and privacy practices. 
%
Compared to the large body of work devoted to understanding and addressing usability issues of crypto wallets, their accessibility is overlooked in prior literature, which we elaborate on in the following section.
\vspace{-2mm}
\subsection{Accessibility of Crypto Wallets}
\vspace{-2mm}
Accessibility in the context of crypto wallets has been discussed and promoted in earlier days. A blockchain workshop position statement highlighted that most crypto wallets relied heavily on visual elements, which made them difficult or even impossible for blind users to use~\cite{unsolved}. 
Advocacy groups, such as the National Federation of the Blind, have responded to the accessibility challenges and called for increased attention to the accessibility of crypto technology. They argued that the decentralized nature of cryptocurrencies and the lack of centralized oversight may hinder accessibility for blind users~\cite{nfb}. However, to date, there has been a lack of academic work on the accessibility of crypto wallets. Thus, it is evident that there is a need for more focused research on the accessibility issues faced by blind users when using crypto wallets, as well as a need for the development of wallet designs that are accessible to all users, including those with disabilities. In this study, we aim to contribute to the limited literature in understanding and addressing accessibility issues of crypto wallets.

\vspace{-2mm}
\section{Empirical Analysis of MetaMask}
\vspace{-2mm}
\label{metamask}
\rev{No prior studies have examined wallet accessibility for blind users. To help fill the gap, we 
first analyzed 10 popular wallets 
in terms of their features and user reviews from three major platforms, i.e., Chrome Web Store (Chrome extension), App Store (iOS), and Google Play Store (Android).
Details of this analysis are in Section~\ref{evaluation} \rev{in the Appendix}. 
Our competitive analysis allowed us to explore potential accessibility challenges which were then used to inform our study and redesign. 
In particular, we found common complaints about the lack of accessibility among these crypto wallets, such as poorly labeled buttons, and learnability and security issues.} 

To complement our aforementioned analysis, we conducted a user study with 10 sighted users, 7 blind users, and one user with low vision with MetaMask (Version 10.23.2). They were recruited from blockchain channels on Discord, Twitter, etc., as well as our participant pools of previous accessibility studies. 
Table~\ref{demographic} in Appendix shows the details of these participants (M1-M18). We followed a similar procedure as in Section~\ref{sec:evaluation} for this exploratory user study, where we asked our participants to conduct a few tasks, such as creating a wallet account and sending some (testnet) tokens. Interviews were conducted before and after the tasks. The whole process generally took 1-2 hours. Blind users spent longer time on the tasks given the accessibility issues. Participants were given \$30 as a compensation. 
Qualitative and quantitative data collected in this stage were used to inform our redesign to improve the accessibility and usability of MetaMask. 

Our data came from participants' think-aloud responses and our observation notes during the tasks as well as the interview responses. The success rate of tasks and results from the SUS survey helped assess usability and accessibility for blind users. The educational aspect was measured through knowledge question (KQ) surveys and task success rates. 
We also examined tasks such as typing the correct receiving address and avoiding seed phrase disclosure to understand usable security implications. 
\vspace{-2mm}
\subsection{Findings: MetaMask}
\vspace{-2mm}
The user evaluations of MetaMask revealed a number of issues about accessibility, security, as well as education about crypto literacy. 
On average, our sighted participants finished 8.6 
of the 10 tasks. Blind users similarly finished 8.3 tasks, but the process was more cumbersome for them. 
It took sighted users 28.2 minutes on average to finish the tasks, while for blind users, the time increased to 47.9 minutes (about 70\% longer than sighted users). A major reason for this time difference was the accessibility challenges encountered by blind users when using the wallet.  
Table~\ref{quantitative:results} summarizes the quantitative results of our evaluations.  
We detail our findings next. 

{\bf Accessibility.} 
The majority of blind 
participants were more or less discouraged by accessibility issues, including unlabeled buttons and other web elements (e.g., input fields), lack of confirmations or notifications, incompatibility with screen readers, and the cumbersome process of dealing with the secret recovery phrase.

Buttons in MetaMask were not properly labeled to be readable by screen readers\footnote{Our participants mostly use JAWS, NVDA, and Voiceover (only available on Mac systems).}, according to many participants such as M13. This made wallet usage awkward for blind users. In the onboarding process, M11 failed to set up his password promptly since the password rule (``8 character min'') was not readable by his screen reader. He also could not reveal the hidden secret recovery phrase in the onboarding process, since the button of ``click here to reveal secret words'' was hard to find and operate with a screen reader; thus he could not go to the next page without the help of our research team. During the transaction process, the field to enter the transaction amount 
was also not labeled (M11, M13), making the transaction a rather time-consuming process for the blind users. 
For some checkboxes, screen readers mistakenly announced them as unchecked even after the users checked them (M18).

Inconsistent notifications were raised as another accessibility issue in MetaMask. Sometimes, announcements were not provided after an operation was performed, e.g., copying the wallet address, or submitting a transaction. On the MetaMask wallet main page, the wallet address is provided. After one hovers over this clickable button with their mouse, a text popup would appear, saying ``Copy to clipboard.'' However, if a blind user navigates to this button with their keyboard, the text will not be verbally announced. Many of our blind participants like M13 did not know how to copy the wallet address until being told by us to press the Enter key on the button. There was also no announcement after the wallet address was copied (M11), while for sighted users, there was a text popup ``Copied!''

When verifying the secret recovery phrase during onboarding or importing accounts, blind users had to spend much time and energy confirming it word by word. In MetaMask, each secret recovery phrase is a random set of 12 words and users need to select the words in the correct sequence to verify it. 
As in M18's case, when confirming the secret recovery phrase, she needed to check the upcoming word in the original phrase, and go through the shuffled words to find the matching one --- this process was cumbersome using a screen reader. M15 also complained that this confirmation process took a lot of energy. 
Some blind users such as M11 and M13 did not want to go back and forth to confirm the secret recovery phrase and chose to use the ``remind me later'' option to skip the process, \rev{which could become a significant security risk as MetaMask did not provide an intuitive way for them to go through this process again later in use.}  
M13 skipped the process and could not import her wallet later since she did not have the secret recovery phrase. M11 downloaded the secret recovery phrase in plain text in a file, which could be easily left in the wrong hands. He felt MetaMask should provide more secure and accessible options for storing the secret recovery phrase.

Many expressed that accessibility issues could also lead to security problems. For example, M11 explained how improper button labeling could lead to security and trust issues for blind users, \textit{``With some of the functionalities unclear to screen reader users, it raises trust concerns in that I'm afraid to set off some unknown function that could negatively impact my account. For example, I might accidentally click a wrong button which is not labeled, and reveal my account. It could be a real problem if I'm on public channels.''} This chain effect of accessibility issues and subsequent fear of accidentally revealing important private information can significantly limit their ability to engage with the wallet and the crypto space. 

While the aforementioned issues were spotted in an earlier version of MetaMask (Version 10.23.2), the more recent version (Version 10.25.0) made some color changes in its light mode to improve color accessibility. However, other accessibility issues we identified still persisted. The design changes we made later were still lacking in the latest version of MetaMask. 
\rev{We were not directly collaborating with MetaMask, but plan to share our findings and redesigns with MetaMask.} 

{\bf Education \& Learnability.}
Except for M3 who was familiar with the concepts of crypto wallets before the study, all other participants found MetaMask confusing due to its many complicated concepts, such as secret recovery phrase, private key, gas fee, and main vs. test networks. 
For instance, many reported a lack of education for gas fees. Some participants, like M6, suspected that the gas fee might be a transaction fee equivalent, but none of them were sure about this concept. M1 was confused about what percentage of the transaction amount she should pay as the gas fee. 
Similarly, many participants did not understand the meaning and importance of the secret recovery phrase even after the study. \rev{These concepts are common in crypto wallets, which could give novice users an extra barrier when using them.}  
M16 thought the option of skipping secret recovery phrase confirmation (``Remind me later'') in the onboarding process diluted the education on this concept, and he was no longer sure if it was important \rev{(``Skippable things are not important'')}. 

MetaMask often failed to provide explanations on crypto concepts, which could be challenging for blind users. For example, the shortened wallet address (e.g., 0x056...8089) was there without further explanation, with many blind participants not knowing it was the wallet address.
Sighted users could infer the role of the string, i.e., they inferred the wallet address string was the address after seeing it: \textit{``I guess it's the wallet address. Addresses usually look like this.''} 
However, blind users found more difficulty doing so since they could not visually see the wallet address to infer what it was. 

Moreover, our participants tended to skip educational videos out of their user habit with apps. Blind users like M16 preferred text-based instructions, since they were easier to read, more accessible, and more time-saving than videos. 
The education provided by MetaMask was associated with a small improvement in the number of KQs answered correctly. The participants answered only 0.3 more questions correctly in the post-study KQ survey than in the pre-study one.

{\bf Usable Security.}
A few participants (both sighted and blind users) mentioned their concerns about sending crypto assets to wrong receiving addresses, since in MetaMask, there was not a confirmation page asking users to double-check their transaction details such as receiving address and amount. 
M9, who manually typed the receiving address during the transaction task, expressed the fear of typing a wrong address. Several blind participants typed their own wallet address instead of the one provided by the research team, and corrected it after being reminded by the research team.


\textbf{Disproportionately Impacted Blind Users.} Our sighted and blind participants experienced security and learnability issues in MetaMask, such as uncertainty about transaction accuracy and a lack of explanation for crypto concepts. These issues were in part because they were novice users of cryptocurrencies and crypto wallets. However, these issues could affect blind users even more. For instance, without accessible (visual) cues, it was harder for blind users to tell different addresses apart. Moreover, inaccessible features such as unlabeled buttons made it a rather cumbersome process to use MetaMask for blind users.   

\vspace{-2mm}
\section{Our Redesign of MetaMask}
\vspace{-2mm}
Previous studies in crypto wallets have primarily focused on investigating the usability challenges~\cite{usability} and security perceptions~\cite{mental_model} of blockchain technologies. 
However, there remains a lack of clarity regarding how to design crypto wallets to accommodate a broader user population, which includes novice users with limited cryptocurrency literacy, and individuals with visual impairments. 
Our competitive analysis of multiple popular wallets (described in Section~\ref{evaluation} \rev{in the Appendix}) and user evaluations of MetaMask (detailed in Section~\ref{metamask}) have revealed several limitations and shortcomings that informed the areas of improvement for accessibility as well as education and usable security. While blind users were disproportionately impacted by the design flaws, we utilized the results to guide our accessibility-centered designs.
To explore ways to address these issues, we employed an iterative design process to implement and evaluate our redesign ideas.

\begin{figure*}[!t]
\centering
\begin{minipage}[b]{0.40\linewidth}
\includegraphics[width=\columnwidth]{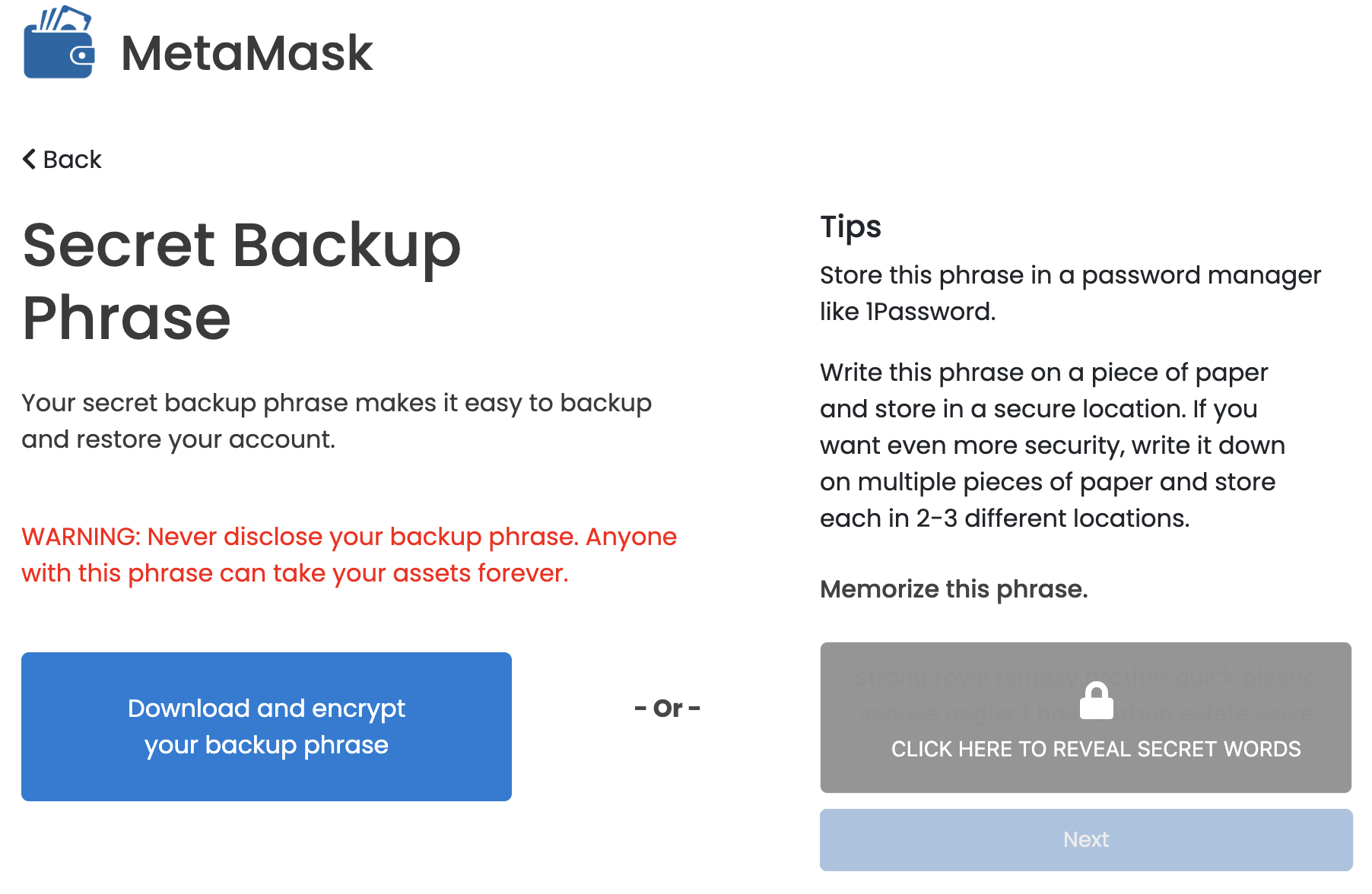}
\caption{Downloadable encrypted secret recovery phrase for seamless management.
}
   \label{fig:fig-ac1}
\end{minipage}
\quad
\begin{minipage}[b]{0.40\linewidth}
\includegraphics[width=\columnwidth]{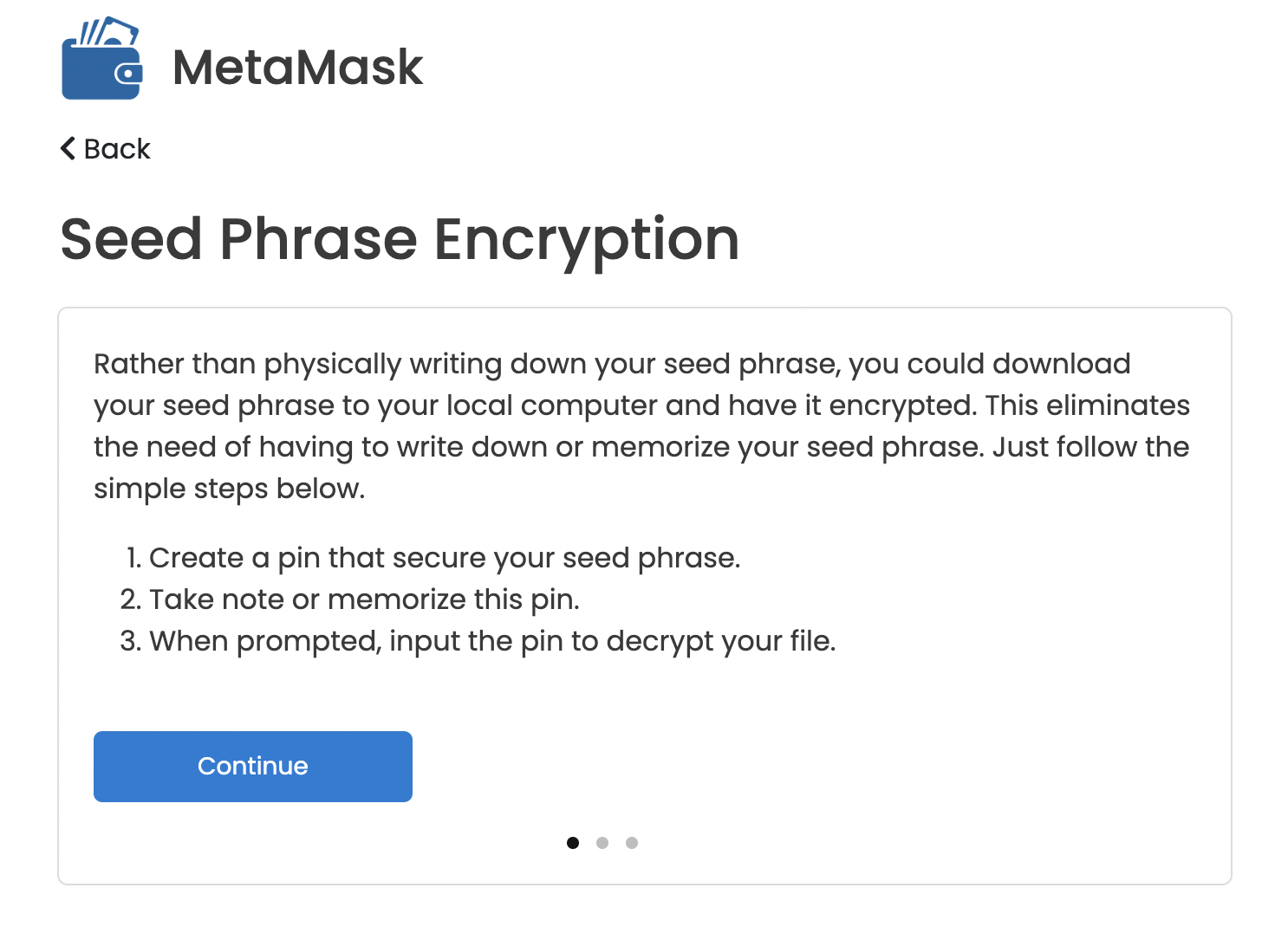}
\caption{A PIN code to encrypt the secret recovery phrase before downloading it to enhance security. 
}
   \label{fig:fig-ac2}
\end{minipage}
\end{figure*}

\vspace{-2mm}
\begin{figure*}[!t]
\centering
\begin{minipage}[b]{0.40\linewidth}
\includegraphics[width=\columnwidth]{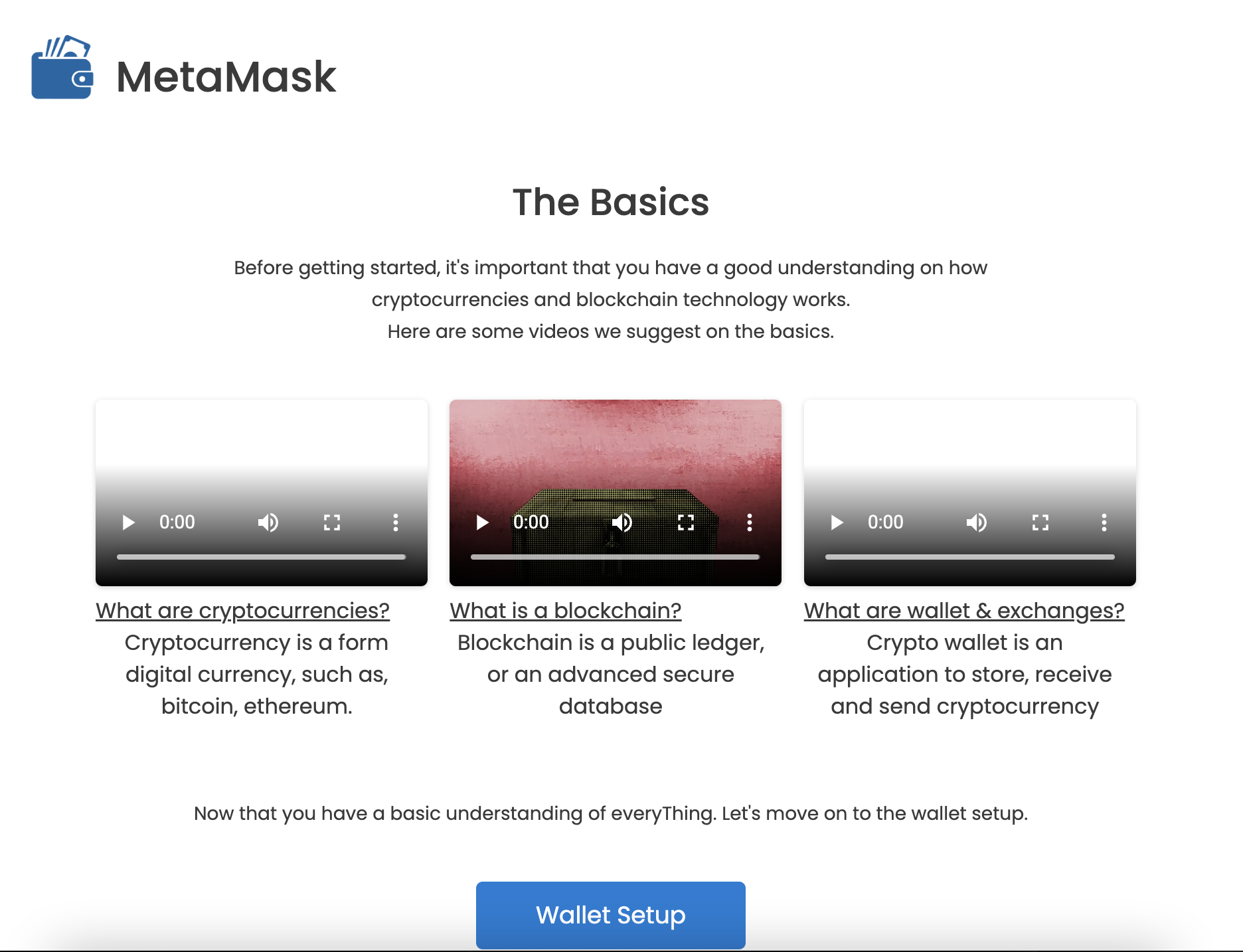}
\caption{A dedicated education page with accessible videos and summative text to educate users on blockchain and wallet basics during onboarding.
}
   \label{fig:fig-edu1}
\end{minipage}
\quad
\begin{minipage}[b]{0.40\linewidth}
\includegraphics[width=\columnwidth]{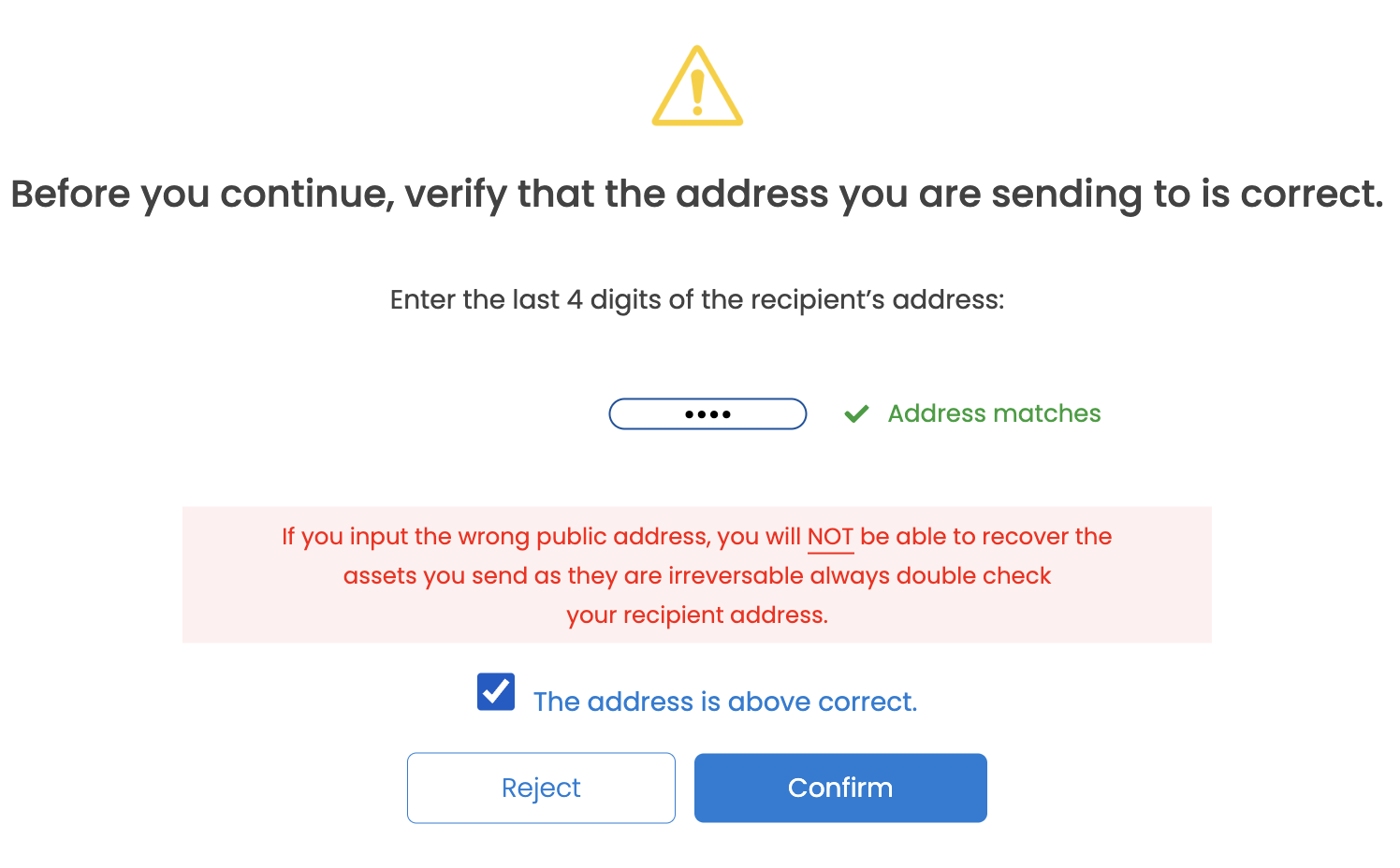}
\caption{A confirmation page for users to double-check the receiving address by re-typing the last 4 digits to improve transaction security. 
}
   \label{fig:fig-us1}
\end{minipage}
\end{figure*}

\vspace{-2mm}

\subsection{Crypto Wallet Redesign Considerations}
\vspace{-2mm}
Existing crypto wallets are often cumbersome to use for blind users due to a number of reasons, such as inadequate labeling, core functions which require great cognitive effort from blind users (e.g., secret recovery phrase management), and ineffective and inaccessible learning resources. 
Thus accessibility became a pivotal design consideration in our study. To enhance accessibility, we aimed to label buttons and other web elements adequately, organize them into a clear hierarchy, and use a combination of colors with sufficient contrast to make the content more distinguishable for users with low vision. 
In addition, we considered streamlining the cumbersome secret recovery phrase management during the processes of onboarding and account importing to improve usability while potentially enhance accessibility for blind users.
\vspace{-2mm}
 \begin{tcolorbox}[width=\linewidth, colback=white!95!black, boxrule=0.5pt, left=2pt,right=2pt,top=1pt,bottom=1pt]
\stepcounter{consideration}
{\bf Design consideration \arabic{consideration}:}
{We improved the accessibility of crypto wallets by labeling buttons adequately, organizing web elements, and simplifying complicated tasks such as secret recovery phrase management.}

\end{tcolorbox}
\vspace{-2mm}

While crypto concepts were found harder to grasp for blind users without visual cues, we aspired to present design features that improved both accessibility and learnability of crypto wallets. To enhance the accessibility of educational resources on crucial concepts and terminologies in cryptocurrency, such as gas fees and secret recovery phrase, we aimed to incorporate intuitive onboarding and transaction processes featuring a well-informed navigation with just-in-time video and text instructions. The embedded instructions and guidance would potentially eliminate the need for users to switch to a separate help center or additional web pages, which could pose a challenge for blind users. The instructional videos and text would educate novice users about crypto concepts such as secret recovery phrase and how to securely use the wallet. 
\vspace{-2mm}
 \begin{tcolorbox}[width=\linewidth, colback=white!95!black, boxrule=0.5pt, left=2pt,right=2pt,top=1pt,bottom=1pt]
\stepcounter{consideration}
{\bf Design consideration \arabic{consideration}:}
{We embedded just-in-time 
 educational resources including videos and text to help blind users understand critical concepts and explore wallets.}

\end{tcolorbox}
\vspace{-2mm}

Finally, we considered incorporating the feature that allowed users to verify the last four digits of the recipient address in the transaction process. This was motivated by the results of our user study with MetaMask, where participants strongly suggested adding a confirmation page when financial assets were at stake. The blind participants often could not tell different wallet addresses apart, which made them prone to the risk of sending crypto assets to the wrong addresses. Such security features could enhance the overall security of users' assets, especially for blind users.
\vspace{-2mm}
 \begin{tcolorbox}[width=\linewidth, colback=white!95!black, boxrule=0.5pt, left=2pt,right=2pt,top=1pt,bottom=1pt]
\stepcounter{consideration}
{\bf Design consideration \arabic{consideration}:}
{We added security features to help reduce risks for novice users especially for blind users.}
\end{tcolorbox}
\vspace{-2mm}
\vspace{-2mm}
\subsection{Redesign: iWallet (V1)}
\vspace{-2mm}


We used MetaMask, the most popular crypto wallet, as the template for our redesign in terms of accessibility, education, and usable security for blind users.

\textbf{Accessibility.}
To improve the accessibility of MetaMask, we took inspiration from Universal Design (UD), a framework for inclusive design, providing principles for accessible experiences in both physical and digital domains~\cite{lima2012analysis}. These principles include equal conditions, flexibility, simplicity, tolerance for error, and reduced physical effort. UD implementation is guided by best practices and accessibility guidelines such as Web Content Accessibility Guidelines (WCAG)~\cite{kelly2009web, wcag} and Americans with Disabilities Act (ADA)~\cite{ada}. More specifically, we first improved labeling for screen readers and darkened texts for better contrast toward meeting the WCAG standards~\cite{wcag}, which MetaMask failed to meet. The ARIA~\cite{aria} specification, \rev{which provides a framework to improve the accessibility and interoperability of web content and applications,} was also implemented for additional contexts for screen readers.
%
To help blind users be aware of their status in operations, we provided notifications in pop-up windows after they finished each task, e.g., submitting a transaction. 
We further designed the function of downloading and uploading an encrypted version of the secret recovery phrase for better accessibility and security for blind users (Figures \ref{fig:fig-ac1} and \ref{fig:fig-ac2}), as the process of writing and typing in the phrase was cumbersome and error-prone for these users.
\


\textbf{Accessible Educational Resources.}
Our redesign in the education aspect utilized two types of media: video and text. Figure~\ref{fig:fig-edu1} displays an example of the redesign aimed at educating users on wallet usage and crypto concepts. Specifically, we added a dedicated page at the beginning of the onboarding process with three informational videos to give users a general understanding of blockchain, cryptocurrency, and crypto wallets/exchanges. Note that the summative text under the videos was implemented in the second design iteration. We also provided a video during the transaction process to explain the concept of gas fee, which was found confusing in previous research~\cite{usability_report} and our evaluation of MetaMask.
Some of our blind participants in the MetaMask evaluation expressed that they skipped videos because they were too fast to grasp, and they would prefer text. 
In response, we added text explanations using metaphors throughout the wallet pages. 
For instance, we used bank account number to explain a wallet address and bank password to explain a private key.  
To educate users about gas fee and its impact on transaction time in an intuitive, direct-manipulation manner, we designed a gas fee slider showing the positive relationship between gas fee and transaction speed, and allowing users to choose between low, average, or high fees for estimated transaction times of 45, 30, or 15 seconds, respectively. 




\textbf{Usable Security.} \label{sec}
To improve transaction security, especially for blind users, we designed a dedicated address confirmation page (Figure~\ref{fig:fig-us1}), asking users to re-type the last four characters of the receiving address; only if there is a match, users can proceed with the transaction. This feature was designed to help ensure crypto assets were sent to the right address. 


\vspace{-2mm}
\subsection{Pilot Study Results of iWallet (V1)}
\vspace{-2mm}
We conducted a formative pilot study with 10 sighted users and 8 blind users (W1-W18) to get feedback about our initial redesign. We followed the same study process as in the evaluation of MetaMask. Main pilot results are summarized below but detailed in Section~\ref{pilot} \rev{in the Appendix}.

The receiving address confirmation was regarded as a useful and accessible security feature. Our embedded education was deemed useful by participants, though the blind users expressed a preference of text, which was more accessible for them, over video instructions. While accessibility was greatly improved in iWallet, leading to lower task completion time than MetaMask, several accessibility challenges remain: (1) Blind users wanted explicit text explanations beside crypto concepts such as wallet address since they could not visually infer their meaning as sighted users; (2) Blind users expected more information (e.g., what is the next page for) in button labeling to assist their navigation; (3) Blind users wanted text summaries to supplement videos so that they could skip the often inaccessible videos without missing important information; (4) We failed to prioritize the option of uploading the secret recovery phrase in the account importing process, leading to task failure. Toward addressing the original accessibility issues and the ones in our initial redesign, we focused on accessibility in the second design iteration, which we elaborate next.

\vspace{-2mm}
\subsection{Redesign: iWallet (V2)}
\vspace{-2mm}

Based on the evaluation of iWallet (V1), our primary design goal was to enhance its accessibility. 
We provided more thoughtful button labeling in a hierarchical manner to mitigate physical and cognitive efforts, and brief descriptions of buttons and web elements to help blind users understand what information was contained in them. We further improved the consistency of page layout and prioritized the download/upload option of managing the secret recovery phrase during account creation and importing, aiming to make this accessibility design better received. 

We also made education more accessible for blind users. To cater to user preferences \rev{and support their accessibility needs}, we added text summaries to supplement educational videos, as some participants indicated that they preferred text instructions, which were more accessible with a screen reader, over videos. For example, under the video on cryptocurrency, we provided a text summary: ``Cryptocurrency is a form of digital currency, such as Bitcoin and \rev{Ethereum (ETH)}.'' Additionally, we made a number of changes to mitigate confusing points of the user interface that our blind users expressed during the pilot study. For example, MetaMask allows for the management of multiple ``accounts,'' i.e., public/private key pairs. By default, MetaMask refers to the wallet address of a user's default account as ``Account 1.'' However, the concept of a wallet address as an ``account'' was confusing to users. While sighted users were able to infer the wallet address string to be an address, blind users found difficulty figuring it out. Thus we changed the heading from ``Account 1'' to ``Wallet Address'' to be more in line with the educational content we showed users on wallet setup. Similarly, we provided a text explanation for ETH, indicating it was a cryptocurrency.



\vspace{-2mm}
\subsection{Wallet System Design and Implementation}
\vspace{-2mm}
We developed our crypto wallet using React JS, and tested it on Chrome and Firefox to ensure its functionality. The backend of the wallet is capable of handling various functions, such as updating users' transaction history when they buy or sell cryptocurrency, facilitating local storage for data requests from APIs, and including general helper algorithms to streamline the redesigned features (e.g., gas slider design, downloadable encrypted seed phrase). Our wallet workflow includes essential features such as account creation, secret recovery phrase encryption, and gas fee adjustment. To save users' transaction activity, our crypto wallet app engine interacts with storage hosted on MongoDB. Whenever users' activity or queries are received, the wallet database updates the state to reflect the current activity on their end. Note that our crypto wallet is an unlisted browser extension, and only selected participants were asked to perform tasks during the study. Our wallet code is available on GitHub\footnote{Wallet code: \url{https://github.com/AccountProject/Wallet_App}}.


\vspace{-2mm}
\section{User Study: iWallet (V2)}
\vspace{-2mm}
\label{sec:evaluation}

In the main user study, we aimed to evaluate our proposed accessibility features after two design iterations. 
Below, we provide a detailed description of the recruitment process, experiment setup, and data analysis methods.
\vspace{-2mm}
\subsection{Recruitment}
\vspace{-2mm}
To evaluate iWallet (V2), we conducted a user experiment with novice blind crypto users (N=8, W19-W26).
We defined novice users as those who may have heard of cryptocurrency but have not traded or used them yet, or those who only had experience with centralized exchanges (CEXes) such as Coinbase and Binance, but had no or little experience with non-custodial wallets such as MetaMask. The screening survey included questions asking about potential participants' experience with cryptocurrency, exchanges, and wallets, as well as demographic information such as gender, age, educational level, and country. Given the focus on accessibility of our wallet (V2), we specifically sought out to recruit blind participants. To this end, we asked about participants' visual acuity and screen reader(s) they used. 

The participants were recruited from various cryptocurrency channels and forums, including Discord, Twitter, and Reddit, as well as previous participant pools of our accessibility studies.
Ultimately, our goal was to recruit a diverse group of participants with varied demographic characteristics, as presented in Table~\ref{demographic} \rev{in the Appendix}. All explicitly indicated an interest in cryptocurrencies and in using crypto wallets.

The participation in our study was completely voluntary, and participants were allowed to withdraw at any time. The participants received \$30, in form of Amazon gift cards, as a compensation. The whole study lasted 1-2 hours for our participants, and blind users tended to spend more time finishing the study. The study was IRB approved. 
\vspace{-2mm}
\subsection{Experiment Setup}
\vspace{-2mm}

\textbf{Procedure.}
We started by conducting brief semi-structured interviews with the participants to gather insights about their prior experience with cryptocurrency and crypto exchanges, if any. After answering six knowledge questions (KQs), which were used to assess their crypto knowledge, they were assigned several tasks regarding crypto assets management and transaction. After finishing the tasks, we asked the participants KQs again to see if our education was effective and well received. We then asked participants to fill in a SUS questionnaire to evaluate the usability of the wallet. Finally, exit interviews were conducted to obtain participants' overall experience of using the wallet, including perceived accessibility and general usability. Suggestions for future wallet design especially regarding accessibility were also collected.
We conducted all user experiments via Zoom, and recorded the interviews and tasks, upon consent, for further analysis. 
Below we detail the study design. 

\textbf{Exploratory Interview.}
After introducing our study to the participants, we asked about their experience and knowledge of cryptocurrency and its underlying infrastructure, i.e., blockchain. Most of our participants were truly novice crypto users, who had only heard of cryptocurrency, and we kept the interviews with them short. For those who had used crypto exchanges before, we asked about their general experiences with exchanges, especially regarding the accessibility aspects. We also asked them what general and accessibility features they would expect if they needed to use a crypto wallet to trade cryptocurrency.

\textbf{Tasks.}
Before tasks began, knowledge questions were asked in a survey to assess participants' initial crypto knowledge, as a reference to that after performing tasks. 
\rev{The six knowledge questions (KQs) were multiple choice questions about core wallet concepts, i.e., token names (ETH), wallet addresses, seed phrases, account security, transactions, and gas fees.} 

Each participant was asked to perform several tasks with our wallet. 
The wallet has been pre-installed on the research team's local computer, and participants were given access to control this computer remotely to perform the tasks, which was a function afforded by Zoom. They were asked to think aloud and answer questions during the study. We recorded the participants' process of performing the tasks for later analysis. 

The tasks (N=4) and sub-tasks (n=9) required to finish each task were revised from ~\cite{usability}, and are listed below.
We specifically sought out to observe how accessible our wallet was to novice/blind users, and identify any accessibility challenges.
\begin{itemize}
    \item \textbf{Task T1}: Configuration - Creating a new account within the wallet (sub-task t1). Participants can optionally watch educational videos to understand the concepts of crypto, blockchain, and wallet. 
    \item \textbf{Task T2}: Checking wallet address, (receiving test ETH), and checking wallet balance. We ask participants to provide their wallet address to us (sub-task t2), and check test ETH balance after receiving it (sub-task t3). 
    \item \textbf{Task T3}: Spend/Transfer - Making a transaction of 1 test ETH to the research team. Thus, this task involves finding the transaction functionality (i.e., Send) (sub-task t4), entering information such as receiver's address provided by the research team (sub-task t5) and ETH amount (sub-task t6), submitting the transaction (sub-task t7), and expressing when they think the transaction is confirmed (sub-task t8).  
    \item \textbf{Task T4}: New device scenario - Imagining using a different device and importing the existing account. We ask participants to refresh the wallet page (get back to the onboarding phase), and ask them to import their existing wallet into the ``new'' device (sub-task t9).
\end{itemize}

After using the wallet, we asked the participants to answer the KQs again to evaluate their crypto knowledge. The SUS survey was filled out to evaluate the usability of the wallet. In addition to the 10 items implemented in the SUS survey, we additionally contained 4 wallet-specific usability questions concerning the onboarding process, wallet address, transaction process, and gas gee.


\textbf{Exit Interviews.}
In the exit interview, we asked participants for their opinions on various aspects of the wallet, including overall experience, accessibility, learnability, security, privacy, etc. We also asked for design suggestions from them to further improve our wallet, especially in terms of accessibility.

\vspace{-2mm}
\subsection{Data Analysis}
\vspace{-2mm}
We performed both qualitative and quantitative analysis on our collected data from interviews, task observations, and surveys. 

\textbf{Qualitative Analysis.}
Two authors independently coded the transcripts of the conversations during the experiments and interviews as well as observational notes, and met regularly to discuss. 
Through thematic coding \cite{thematic}, themes started to merge and brought us back to the transcripts to find more data for them. We used XMind~\cite{xmind}, a mind mapping tool, to arrange and organize codes and corresponding quotes into a hierarchy of themes. After several iterations of analysis, we arrived at the current findings. We use observational notes and participant quotes to illustrate our points. All quotes have been anonymized to protect privacy of the participants. 

\textbf{Quantitative Analysis.} 
Recorded videos of the participants performing assigned tasks were analyzed to measure task success rates and task completion times. Surveys, including two KQ surveys and one SUS questionnaire, were statistically analyzed to understand educational effect and user experience of the wallet. 




\vspace{-2mm}
\section{Findings: iWallet (V2)}

\label{sec:findings}



\vspace{-2mm}
\rev{
\begin{table*}[h]
\footnotesize
\centering
  \begin{tabular}{ccccc}
    \toprule
    Wallet  &
    SUS score (out of 100) & 
    \begin{tabular}[c]{@{}c@{}}Task success rate (\%)\end{tabular} & 
    \begin{tabular}[c]{@{}c@{}}Task completion time (min)\end{tabular} & 
    \begin{tabular}[c]{@{}c@{}}Improvement in KQs \end{tabular} \\
    \midrule
    MetaMask & \rev{70} & 86 & 28.2 (sighted), 47.9 (blind) & +0.3 \\
    \midrule
    iWallet (V1) & \rev{65} & 79 & 26.1 (sighted), 40.0 (blind) & +0.7 \\
    \midrule
    iWallet (V2) & \rev{\textbf{81}} & \textbf{100} & \textbf{37.8 (blind)} & \textbf{+1.3} \\
    \bottomrule
  \end{tabular}
  \caption{Quantitative results of our evaluations.}~\label{quantitative:results}
\end{table*}
}
\vspace{-2mm}

Our accessibility features were greatly improved compared to MetaMask and the previous version of iWallet (V1). \rev{iWallet (V2) outperformed MetaMask in terms of SUS score, task success rate, task completion time for blind users, and improvement in the number of KQs answered correctly after usage (see Table~\ref{quantitative:results}).}  
The security and education features also helped blind users interact with crypto wallets smoothly and safely.  Our participants provided rich qualitative insights regarding our accessibility and security improvement.

\vspace{-2mm}
\subsection{Accessibility}
\vspace{-2mm}
Our improvement in accessibility resulted in a higher SUS score \rev{(81)}, compared to \rev{70} for MetaMask and \rev{65} for our previous iteration. Moreover, none of our participants encountered any difficulties in completing the sub-tasks. Our blind participants spent 37.8 minutes on the tasks, compared to 40 minutes in V1 and 47.9 minutes in MetaMask.

{\bf Enhancing Accessibility for Equitable Use.} Accessibility was key to blind users' interaction with crypto wallets and financial apps in general, but was frequently overlooked. Our participant, W26, reported her experience of trying multiple centralized and decentralized crypto exchanges before settling down to Robinhood, which worked well with iPhone and Voiceover. She added that Coinbase, another crypto exchange, was not accessible, leading to a lack of confidence in using the platform: \textit{``It's confusing, not very friendly. I cannot get through it with confidence. If I accidentally hit a button When dealing with finances, I don't know if I will get into an area that I cannot get out of. Buttons should be labeled correctly instead of just reading `button' or `link'.''} 
After using our wallet, she thought it well met her accessibility needs, making it possible and enjoyable for her to interact with crypto wallets. Many of them did not expect crypto wallets to be accessible, and felt it a pleasurable process to use our wallet. W20 expressed appreciation for our accessible wallet and inquired about its availability in the Chrome Web Store and Apple App Store for daily use. 

{\bf Adequately Labeled Buttons for Intuitiveness.} Our participants provided positive feedback regarding the clear and informative button labeling (W20, W21, W22, W23, W25, W26). Participant W23 highlighted the usefulness of the descriptive button labeling in guiding the completion of tasks, stating, \textit{``I found the button names to be descriptive compared to other apps I usually use. When I complete a task and press the next button, I'm quite sure what I'm going to see next, like, the next is X or Y.''}
W25 similarly praised the well-labeled buttons: \textit{``The buttons and input fields are pretty well labeled.''} 

{\bf Onboarding/Portability with Low Physical Effort.} Our participants, e.g., W21, indicated experiencing no accessibility issues when making transactions and found the user interface workflow to be straightforward. W19, W20, W21, W23, and W26 found the streamlined onboarding process -- whether creating a new wallet or importing an existing one -- to be effortless. W19 expressed that \textit{``the onboarding process is pretty easy to follow,''} while W23 echoed her opinion, \textit{``Account creation is intuitive and the steps are straightforward.''} W26 was particularly intrigued by the onboarding process, which introduced her to the new concept of secret recovery phrase -- something she had never encountered in her previous experience with Robinhood. 
Most participants preferred the option to download the encrypted secret recovery phrase over the option to write it down manually (W20, W22, W23, W24, W25, W26). 
W22 opted to upload the secret recovery phrase to verify the account, explaining that \textit{``selecting the words takes too much time, and uploading seems less time-consuming.''} 
W20, who chose to write down the phrase, had difficulties importing his wallet due to not inserting space between words while typing them manually. He stated that the process would have been easier if he had chosen the option to download the phrase.
After \rev{the researchers} consistently prioritized the download/upload option in the onboarding/importing process, none of the participants in this round had confusion as observed in the pilot study.

{\bf Enhancing Accessibility Enhances Security.} The downloading option of managing the secret recovery phrase was also regarded as more secure since it was encrypted by a pin (W24). W23 further expressed his security concern of the writing option: \textit{``If we are outside, there would be privacy issues. When we write it down in a mobile phone or computer, other people can hear it. Not all screen reader users use headphones.''} 
Interestingly, several participants downloaded and wrote down the secret recovery phrase at the same time to avoid losing it (W21, W22, W24). This could be due to blind users' typically cautious behavior when using a new app.

The only remaining accessibility improvement recommendation pertained to ensuring seamless auto-focus on popup windows for screen reader users (W22, W24, W25, W26). According to W25, who used JAWS as the screen reader, the confirmation notification after the transaction was helpful for blind users; however, the appearance of popups was not spotted by the screen reader timely. As a result, she had to navigate to the bottom of the main page before finding the popup window. W24 suggested using dialog boxes rather than pop-ups to facilitate easier navigation for blind users.

\vspace{-2mm}
\subsection{Accessible Crypto Wallet Education}
\vspace{-2mm}
It was commonly acknowledged by our participants that having sufficient knowledge of crypto wallet was crucial for blind users since they found difficulty relying on visual cues to infer meanings of new concepts and inform their security decisions, echoing with our previous user evaluations. Before the tasks, most of our participants did not know basic concepts in crypto wallets such as secret recovery phrase and wallet address, as shown in the first KQ survey. W24 explicitly indicated that she was \textit{``not sure what secret recovery phrase and wallet address meant.''} W19 was on-boarded to a crypto wallet by her brother, but did not use it ever, as she found it conceptually harder to use than traditional financial apps. She added, \textit{``Blockchain concepts and terminologies are a totally different world.''} 
Before the experiments, our participants answered 3.3 questions correctly out of a total of 6 on average. Following the experiments, the number increased to 4.6.  
\rev{After using our wallet, 10 more participants (from 6 to 16) correctly answered the question on wallet address, compared to an increase of two for the baseline MetaMask.}  
Such evidence demonstrates the efficacy of our accessible education designs, including rich instructional videos and concise text summaries as suggested by blind users in the pilot study, as well as text explanations of wallet address, ETH, etc.

{\bf Crypto Knowledge.} The instructional videos were effective in familiarizing some users with crypto concepts (W21, W22, W23, W25, W26). W22 found the videos useful for learning how to create and use a new wallet: \textit{``The information was totally new for me. That's why I watched the videos.''} W25 intended to watch all the videos 
to \textit{``make sure what it is, in case doing something incorrectly.''} She thought the videos were concise and taught her interesting concepts such as mining. W26 thought the videos were of good length and helped explain things, such as what secret recovery phrase really meant. 
W24 found the videos easy to understand, played at reasonable speed, jargon-free, and informative. On the contrary, W20 did not watch any videos in the onboarding process and had difficulty understanding crypto terminologies such as the secret recovery phrase in subsequent steps. He acknowledged that he would have watched the videos if he knew there were related operations afterward. Additionally, some participants, such as W24, preferred to read the text summaries of the video content rather than watch the videos themselves. 

By explicitly explaining the ``Wallet Address'' right above the string and putting the text ``ETH is a digital currency'' below the ETH balance, we enabled our participants to locate them promptly (W19, W20, W24). In contrast, there were no such text cues in MetaMask and our previous version, making it harder for blind users to infer their meanings. Many of our participants, including W19, indicated using the explicit text evidence to find the wallet address. 

{\bf Goal-Directed \& Engaging Learning.} The gas fee slider was deemed as an intuitive way to adjust gas fee by our participants (W19, W20, W23). W20 thought it was a nice design to have in the wallet. He further explained that he would select the low option when sending money to friends and the high option if he was sending money for business or critical transactions to make it more timely. 
W25 went with the default gas fee, since she \textit{``didn't know enough about it [gas fee] to make it go through.''} W22 also chose the default gas fee because she thought \textit{``the system default would be a good option, and users tend to think the default one is the best option for them.''} W24 further suggested a design to show how many people were choosing each option to inform new users' decisions and increase their trust of the slider as a social navigation feature.

\vspace{-2mm}
\subsection{Usable Security Features}
\vspace{-2mm}
Our participants anticipated a wide range of security measures other than passwords in a crypto wallet in the pre-experiment interview, including pincode (W19), two-factor authentication such as Google Authenticator (W24, W25), account/password recovery, e.g., when losing the initial device (W25), and biometric authentication, such as fingerprints and face recognition (W19, W24), especially in the transaction process where stakes became higher. 
After the study sessions, all participants felt iWallet was secure. For instance, W25 thought iWallet was safe since \textit{``the secret recovery phrase is encrypted [into the downloaded file] and cannot be read by others.''} 
W22 both downloaded and manually wrote down the secret recovery phrase, and thought it as two-factor authentication. W23 recalled the instructional videos reminded him about password security: \textit{``[It] should be strong, not a human or pet name.'' }

The security feature of asking users to re-type and check the last 4 characters of the destination address was well received by our blind participants (W21, W22, W25, W26), who could not easily notice different addresses. W21 interpreted the feature as a way to verify who to send money to, which was a common understanding among them. W25 added that in regular apps, when people made transactions, there would be prompts like \textit{``are you sure it's the right person?''} W22, who reconfirmed the last 4 characters multiple times, thought it as a very useful option. However, one participant (W25) noted that it could be hard to catch the last 4 characters with a screen reader and retype them in a single pass.
\vspace{-2mm}
\section{Discussion}
\vspace{-2mm}
\label{sec:discussion}

In our study, we utilized a competitive analysis of existing wallets, semi-structured interviews, and task-based experiments to identify major accessibility issues experienced by users which led to both learnability and security challenges for blind users. We further implemented and evaluated accessibility-centered design solutions for crypto wallets. Our findings shed light on addressing the accessibility, security, and learnability challenges faced by blind users when using crypto wallets.

In this section, we reflect on accessibility issues in the emerging application domain of crypto wallets and how we alleviated some of these challenges through an iterative design approach. In addition, we discuss education, usable design, and security interventions for improving accessibility. Our goal is to contribute to empirical knowledge in accessibility issues associated with crypto wallets; particularly the intersection of accessibility and security for blind users~\cite{napoli2021m, ahmed2015privacy}. 
\vspace{-2mm}
\subsection{Accessibility Helps Usable Security}
\vspace{-2mm}
Accessibility is of utmost importance for security for blind users.
Many user-facing security concerns in the crypto space stem from user misconceptions \cite{mental_model, mai2020user}. Inaccessible wallets leave blind users especially prone to these misconceptions, and in turn, vulnerable to security breaches. We extend prior work~\cite{napoli2021m} by identifying several instances where security information and consequences are not effectively communicated to users via design and assistive technology in the context of crypto wallets. Participants in our study identified ill-fitting security and crypto concepts on MetaMask, making it difficult for them to 
internalize the consequences and 
make informed decisions during tasks. For example, seed phrase was not conceptually perceptible for blind users to understand the length of security exploitation if they did not save it securely. In addition, MetaMask provided little audio guidance for web elements, misleading screen reader users during use.  

To address these design issues, iWallet incorporated redesigned features that better served blind users in transitioning towards more beneficial states of security awareness. For example, we implement a feature prompting
users to confirm/re-type the receiving address during transactions, which helps blind users ensure they are sending crypto to the correct addresses, and 
potentially helps them avoid ``clipboard hijacker'' which replaces crypto wallet addresses with lookalikes~\cite{pillai2019smart}. This design was especially praised by our blind participants since they could not visually differentiate between different wallet addresses like sighted users and were thus more prone to accidentally sending assets to the wrong addresses. MetaMask did not require address confirmation, and several blind users sent assets to their own addresses instead of the one provided by the research team. In short, by improving accessibility, we also improved security.

\rev{Another key challenge of crypto wallets for our novice user participants was the lack of fundamental knowledge about cryptocurrencies and wallets \cite{energy}. Therefore, we incorporated educational materials about crypto concepts and security into our wallet redesign. We emphasize the importance of learning these critical yet hard-to-grasp concepts by closely collaborating with blind users through the iterative design process. 
Technical concepts are difficult for novice users to grasp. Accessible designs such as textual explanations about these concepts should be included which blind users can access and benefit from. These accessible educational materials can contribute to the usable security for blind crypto wallet users.} 

\vspace{-2mm}
\subsection{Improving Crypto Wallet Accessibility}
\vspace{-2mm}
Though accessibility issues in crypto wallets are well known \cite{unsolved}, little effort has been devoted to understanding and improving them. In our user testing of MetaMask, we found that blind users encountered numerous unlabeled buttons, rendering them unable to complete critical steps, including revealing secret recovery phrases, entering transaction amounts, etc. Some participants skipped the secret recovery phrase backup process during onboarding due to the frustrating accessibility, which could potentially lead to severe security risks. Our redesign and evaluation indicated that adhering to accessibility standards and best practices such as WCAG~\cite{wcag} and ARIA~\cite{aria} could address some critical accessibility issues (e.g., unlabeled buttons) and foster greater adoption of crypto wallets. Future design of crypto wallets should explicitly consider accessibility and implement accessibility best practices.

Moreover, blind users have their unique challenges and needs than sighted users. Visual cues may be quickly grasped by sighted users. For example, after seeing the wallet address string, they immediately knew it was the wallet address without explicit explanations. However, blind users could hardly use such visual evidence to infer unfamiliar concepts. It is also harder for blind users to infer the status of an operation, e.g., if a transaction is confirmed, while sighted users find less difficulty spotting content change on the page, e.g., the change of ETH balance. To improve accessibility for blind users, it is essential to provide more explicit explanations and notifications that can boost their confidence in using crypto wallets. While our notifications in form of popup windows are not accessible enough for blind users, we recommend providing sound notifications which are clear and distinguishable \cite{sound}. For example, different sounds can be used for different types of notifications, such as new transactions or errors. 

Similarly, we observed that some simple operations for sighted users were conceptually or practically harder for blind users. For example, confirming the secret recovery phrase by re-arranging the shuffled 12 words into the original order was extremely hard with a screen reader. The blind users had to go back and forth to check each word and select them in MetaMask. Correspondingly, we designed an additional option for users to download and upload the secret recovery phrase seamlessly, which was well received by them, finding it time-saving and convenient. The encryption feature further added to the perceived security of the wallet.

Through our iterative design and evaluation, we are aware that adhering to accessibility standards and guidelines is important, but not enough. 
Application context is crucial for accessibility development. In financial scenarios, blind users expect more notifications and accessible security measures to allow them to navigate the pages with confidence. Thus crypto developers, who are often sighted, should consider accessibility as a key design requirement for their wallet design. 
\rev{Both accessible content features (e.g., educational videos, screen reader compatible text descriptions) and design features (e.g., gas fee slider, downloadable encrypted secret recovery phrase, receiving address checking) could be helpful for blind users.}






\vspace{-2mm}
\subsection{Limitations and Future Work}
\vspace{-2mm}
There are a few limitations of our presented work. \rev{First, our user studies and redesigns only focused on MetaMask and thus we can not claim that the set of crypto wallet accessibility issues we found was exhaustive. However, these issues were common in other wallets as shown in our competitive analysis.} 
Second, while our sample size is on par with or even larger than that of many other usable security studies with blind users, a larger sample of blind users with diverse (technical) backgrounds would be useful. 
Third, our tasks may not fully simulate and reveal users' actual behaviors in real transactions when stakes become higher (e.g., transactions involving their own real crypto assets). Future work could deploy the crypto wallet and study user behavior in practice. 
\rev{Last but not least, there are limitations of our accessibility designs. For example, encrypting the secret recovery phrase with a PIN means that users need to safeguard and recall the PIN, which could also become a target for attackers. Future work could consider addressing these limitations and proposing additional designs, e.g., security mechanisms that completely remove the need for having the secret recovery phrase.}
\vspace{-2mm}
\section{Conclusion}
\vspace{-2mm}
\label{sec:conclusion}

We presented an iterative redesign of MetaMask to make it more accessible, usable, and secure for novice, blind users. Building on the perspectives of 23 blind users, one low-vision user, and an additional 20 sighted users (N=44), we uncovered a number of accessibility problems with MetaMask that frustrate blind users and that leave them disproportionately prone to common security risks. Our summative evaluation suggests that our redesign alleviated many of these problems: e.g., we followed best practices for accessible design to allow screen readers to access the user interface elements, presented an option for downloading one's secret recovery phrase in an encrypted file to circumvent the need to manually write it down, and created accessible text- and video-based guides to help users understand crypto concepts. Building on these findings, we proposed design implications for creating a more accessible and secure crypto infrastructure for blind users.

\bibliographystyle{plain}
\bibliography{usenix2022_SOUPS}

\appendix
\label{appendix}

\section{Evaluation and User Review of Wallets}
\label{evaluation}
\begin{table*}[t]
\footnotesize
\centering
\begin{tabular}{|p{1.5cm}|p{1.2cm}|p{1.2cm}|p{1.2cm}|p{1.2cm}|p{1.2cm}|p{1.2cm}|p{1.2cm}|p{1.5cm}|}
    
     \toprule
    {\small\textit{Wallet}}
     & {\small \textit{Multiple Platform}}
    & {\small \textit{Seed Phrase Security}}
       & {\small \textit{Multi-sig Security}}
    & {\small \textit{Two-factor Auth}}
    & {\small \textit{Education Page}}
     & {\small \textit{Gas/Traffic Meter}}
       & {\small \textit{Dapp Integration}}
          & {\small \textit{Accessibility}}
       \\
 
    \midrule
     MetaMask & \cmark&\cmark& \xmark& \xmark&\xmark&\xmark&\cmark&\cmark\\
     MyEtherWallet &  \cmark& \cmark& \xmark& \xmark&\cmark&\xmark&\xmark&\xmark\\
     TrustWallet & \xmark & \cmark&\xmark& \xmark&\cmark&\xmark&\cmark&\xmark\\
     Coinbase &  \cmark& \cmark&\xmark& \xmark&\xmark&\xmark&\cmark&\cmark\\
     Exodus&  \cmark & \cmark&\xmark& \xmark&\cmark&\xmark&\xmark&\xmark\\
     Argent &\xmark &\xmark & \cmark& \xmark&\xmark&\xmark&\cmark&\xmark\\
     Jaxx &  \cmark& \cmark& \xmark& \xmark&\xmark&\xmark&\cmark&\xmark\\
     DeFi&\xmark  &\cmark& \xmark& \cmark&\cmark&\cmark&\xmark&\xmark\\
     Lumi&\xmark  &\cmark& \xmark& \xmark&\xmark&\xmark&\xmark&\xmark\\
     Atomic&  \cmark & \cmark&\xmark& \xmark&\cmark&\xmark&\xmark&\xmark\\

    \bottomrule
  \end{tabular}
  \caption{Competitive analysis of existing crypto wallets.}~\label{comp-analysis}
\end{table*}

With our goal of improving usability, before initial designs, we got familiar with and systematically evaluated 10 popular crypto wallets on Ethereum, in 3 aspects, i.e., education, security, and accessibility. Since some wallets were developed and used in multiple platforms (i.e., mobile, browser, and desktop), we evaluated them independently, and analyzed the consistency across different platforms. These 10 wallets are: MetaMask (Browser \& Mobile), MyEtherWallet (Browser \& Mobile), Trust (Mobile), Coinbase Wallet (Mobile), Exodus (Desktop \& Mobile), Argent (Mobile), Jaxx Wallet (Desktop \& Mobile), DeFi Wallet (Mobile), Lumi (Mobile), and Atomic (Desktop \& Mobile). We also analyzed user reviews from various sources such as Chrome Web Store (chrome extension), App Store (iOS), and Google Play Store (Android) to gain additional insights on each wallet.




\textbf{Evaluation}
\textit{General Features.} According to our evaluation, we found that the wallets contained similar functions and processes: an onboarding process where users got and were asked to keep their seed phrase, which was the only way to access their wallets, buying/sending/receiving/exchanging crypto assets, and displaying chart information about each asset's market values and trends. Table~\ref{comp-analysis} lists features in different wallets.

Cross-platform wallets allowed users to access their assets on different devices and with different operating systems. However, only half of the wallets evaluated had an alternate platform besides mobile. This could be an accessibility concern for those who wanted to use a crypto wallet but did not have a smart phone.


A notable feature comes from the DeFi mobile wallet, which integrates a status meter for users to check on current gas prices,
with estimated transaction times, cryptocurrency values, USD values, and graphics for showing how much traffic is currently on the Ethereum network. This feature may help users make better-informed financial decisions and develop a greater awareness of how gas prices work. Our later design of the gas fee slider has been inspired by this feature.

\textit{Education.} A main usability issue is the lack of educational resources for users. Half of the wallets contained education resources and features, but typically only provided one or two screens. MyEtherWallet (mobile) was the only wallet that had a dedicated resource page, educating users on cryptocurrency, blockchain technology, security/privacy best practices, and the wallet itself. However, these resources were not properly embedded in the browser version of MyEtherWallet: the resources were externally provided, and one must log out of their wallet to access them.

MetaMask (mobile) was the only wallet that provided a tutorial for users, teaching them about different parts of the app and what they could do. However, this tutorial was only accessible as a part of the onboarding process, with no other way of accessing it in the future or daily use. This tutorial was also completely absent from its browser extension counterpart, again displaying cross-platform inconsistency.

Several wallets included tips or popup screens that appeared after clicking a ``help'' icon, especially for certain screens/features which required additional clarification. But they were not present in other screens where a user may need help or instructions. For example, MetaMask (mobile), Argent, and DeFi only provided instruction in the settings section, specifying their security measures. The remaining screens were designed with the assumption that users understood their content.

\textit{Security.} Out of the 10 wallets evaluated, nine used a mnemonic/seed phrase for account backup and recovery. The only wallet without a seed phrase mechanism, Argent, alternatively utilized a multi-signature authentication for the same purpose. Specifically, one could choose trusted people as ``guardians'' to help recover the wallet and approve transactions. Other common security measures in these wallets included daily transaction limits, auto-locking, and 2-factor authentication via Web2 intermediaries such as phone/email or a third-party authenticator.

Some wallets, such as MetaMask chrome extension, does little to guarantee security/accuracy of transactions. Specifically, it did not include a confirmation mechanism to verify receiving address during performing transaction. As a result, users would miss the opportunity to double check, which may lead to sending crypto to a wrong address. 

\textit{Accessibility.} During the wallet accessibility evaluation, we used a contrast checker \cite{aim} to check the color contrast of MetaMask as well as Coinbase wallet. In MetaMask, the contrast ratio between foreground color (i.e., text) and background color was $4.27:1$, which did not meet the standard of WCAG 2.0 level AA, which is a standard to assess web accessibility compliance \cite{accessibility:standard}.

\textbf{User Reviews}
We analyzed users reviews to gain additional insights on each wallet, and identify features/functions users wanted. To have an abundant collection, we collected reviews for each wallet from Apple App Store (for iOS version), Google Play Store (for Android version), and browser stores of Chrome and FireFox (for browser extension version). From the thematic analysis of the user reviews, we similarly found three main areas of improvement, i.e., education, security, and accessibility. Here we present representative issues encountered by users, as well as design suggestions and preferences provided by them.

\textit{Education - MetaMask [Firefox Browser Add-on]} 
Users frequently indicated the need for educational resources and instructions during the onboarding process of crypto wallets. One such example was from a MetaMask user who complained about the lack of (jargon-free) instructions in MetaMask: \textit{``I'm a newby with little experience. The instructions from MetaMask are confusing and often lead to screens where there are no direct instructions on what to do next. MetaMask seems to lack personnel who can write instructions in a clear (without jargon) style.'' - Firefox user 16749602, 1 star (03/12/2021)''}

\textit{Security - MetaMask [Google Play Store]}
The limited usability and security afforded by seed phrase is repeatedly complained about. Some users mentioned the current process of dealing with the seed phrase was time consuming, and less usable than the traditional security measures they usually used in Web2, such as Google Authenticator: \textit{``I think there could be more security options, like google authenticator, email verification! A friend of mine was hacked last week, by some sort they got possession of his 12 word seed phrase and boom 5k lost!'' - Pedro Santos, 4 stars (05/07/2021)}

\textit{Accessibility - MyEtherWallet [App Store]}
Yet another important area of improvement is accessibility for engaging a broader audience to the crypto space. Regarding accessibility, many users encountered unlabled buttons and input fields while using crypto wallets, e.g., in the following review,
    \textit{``Buttons need to be labeled for TalkBack users who are blind. I'm unable to create or sign in to the app, because a fair number of the buttons are not labeled so I can't tell if I'm entering in something correctly or not let alone if it is correct and heading the correct or wrong button.'' - chuck winstead, 1 star (03/23/2021)}


The above analysis of wallet features and user reviews has heavily impacted our redesign process. 

\section{Participant Demographics}
We strove to recruit a diverse set of participants for our user evaluations, in terms of age, gender, profession, and vision ability. By so doing, we aimed to identify unique needs of potential crypto users, especially those who were blind, and inform inclusive designs for our crypto wallet. \rev{Most participants did not have prior experiences with cryptocurrencies before the study. Only six out of 44 had some prior experience with centralized exchanges such as Binance and Coinbase.} More demographic details of our participants are summarized in Table~\ref{demographic}.

\begin{table*}[h]
\footnotesize
\centering
  \begin{tabular}{m{10mm}<{\centering}m{12mm}<{\centering}m{15mm}<{\centering}p{10mm}<{\centering}m{10mm}<{\centering}m{35mm}<{\centering}m{10mm}<{\centering}m{15mm}<{\centering}}
    \toprule
    ID  &
    \begin{tabular}[c]{@{}l@{}}Platform\end{tabular} & 
    \begin{tabular}[c]{@{}l@{}}Country\end{tabular} & 
    \begin{tabular}[c]{@{}l@{}}Gender\end{tabular} & 
    \begin{tabular}[c]{@{}l@{}}Age\\ Group\end{tabular} & 
    \begin{tabular}[c]{@{}l@{}}Occupation\end{tabular} &
    \begin{tabular}[c]{@{}l@{}}Crypto\\ Experience\end{tabular} &
    \begin{tabular}[c]{@{}l@{}}Visual \\ Impairments\end{tabular}\\
    \midrule
    M1 & MetaMask & Germany & Female & 25-34 & PhD Student & N & No \\
    M2 & MetaMask & China & Female & 18-25 & Master Student & Y & No \\
    M3 & MetaMask & China & Female & 25-34 & Master Student & Y & No \\
    M4 & MetaMask & Sweden & Female & 18-25 & Master Student & N & No \\
    M5 & MetaMask & US & Female & 18-25 & PhD Student & N & No \\
    M6 & MetaMask & China & Male & 18-25 & Data Scientist & N & No \\
    M7 & MetaMask & US & Male & 18-25 & Self-employed & N & No \\
    M8 & MetaMask & Netherlands & Male & 25-34 & Vehicle Engineer & N & No \\
    M9 & MetaMask & India & Male & 25-34 & PhD Student & Y & No \\
    M10 & MetaMask & Switzerland & Male & 18-25 & Administration & N & No \\
    \rowcolor{mygray}
    M11 & MetaMask & US & Male & 45-54 & Accessibility Expert & Y & Blind \\
    \rowcolor{mygray}
    M12 & MetaMask & US & Female & 35-44 & Logistics Management Specialist & N & Blind \\
    \rowcolor{mygray}
    M13 & MetaMask & US & Female & 45-54 & Contractor & N & Blind \\
    \rowcolor{mygray}
    M14 & MetaMask & US & Male & 25-34 & PhD Student & N & Blind \\
    \rowcolor{mygray}
    M15 & MetaMask & US & Agender & 35-44 & Freelance Artist & N & Low-vision \\
    \rowcolor{mygray}
    M16 & MetaMask & US & Male & 25-34 & Master Student & N & Blind \\
    \rowcolor{mygray}
    M17 & MetaMask & US & Male & 45-54 & Financial Consultant & N & Blind \\
    \rowcolor{mygray}
    M18 & MetaMask & US & Female & 25-34 & Self-taught Student & N & Blind \\
    \midrule
    W1 & iWallet(V1) & US & Male & 18-25 & Master Student & N & No \\
    W2 & iWallet(V1) & US & Female & 25-34 & PhD Student & N & No \\
    W3 & iWallet(V1) & Spain & Female & 18-25 & Translator & N & No \\
    W4 & iWallet(V1) & US & Male & 18-25 & Store Worker & N & No \\
    W5 & iWallet(V1) & Netherlands & Male & 18-25 & Journalist & N & No \\
    W6 & iWallet(V1) & Nigeria & Female & 18-25 & Undergrad Student & N & No \\
    W7 & iWallet(V1) & China & Female & 18-25 & Accountant & N & No \\
    W8 & iWallet(V1) & Hong Kong & Male & 18-25 & Master Student & Y & No \\
    W9 & iWallet(V1) & China & Female & 18-25 & Project Manager & N & No \\
    W10 & iWallet(V1) & US & Male & 25-34 & PhD Student & N & No \\
    \rowcolor{mygray}
    W11 & iWallet(V1) & US & Female & 25-34 & Unemployed & N & Blind \\
    \rowcolor{mygray}    
    W12 & iWallet(V1) & US & Female & 55-64 & Accessibility Specialist & N & Blind \\
    \rowcolor{mygray}
    W13 & iWallet(V1) & US & Male & 25-34 & Accessibility Evangelist & N & Blind \\
    \rowcolor{mygray}
    W14 & iWallet(V1) & Italy & Female & 35-44 & PhD Student & N & Blind \\
    \rowcolor{mygray}
    W15 & iWallet(V1) & US & Male & 18-25 & Undergrad Student & N & Blind \\
    \rowcolor{mygray}
    W16 & iWallet(V1) & Bahrain & Male & 25-34 & Administration & N & Blind \\
    \rowcolor{mygray}
    W17 & iWallet(V1) & UK & Male & 35-44 & Open-source Developer & N & Blind \\
    \rowcolor{mygray}
    W18 & iWallet(V1) & Canada & Male & 25-34 & Developer & N & Blind \\
    \rowcolor{mygray}
    \midrule
    \rowcolor{mygray}
    W19 & iWallet(V2) & US & Female & 35-44 & Accessibility Consultant & N & Blind \\
    \rowcolor{mygray}    
    W20 & iWallet(V2) & US & Male & 18-25 & Lab Scientist & N & Blind \\
    \rowcolor{mygray}
    W21 & iWallet(V2) & US & Male & 18-25 & Therapist & N & Blind \\
    \rowcolor{mygray}
    W22 & iWallet(V2) & India & Female & 35-44 & Bank Manager & N & Blind \\
    \rowcolor{mygray}
    W23 & iWallet(V2) & India & Male & 25-34 & Partnership Specialist & N & Blind \\
    \rowcolor{mygray}
    W24 & iWallet(V2) & UAE & Female & 18-25 & High School Graduate & N & Blind \\
    \rowcolor{mygray}
    W25 & iWallet(V2) & US & Female & 45-54 & College Student & N & Blind \\
    \rowcolor{mygray}
    W26 & iWallet(V2) & US & Female & 55-64 & Self-employed & Y & Blind \\
    \rowcolor{mygray}
    \bottomrule
  \end{tabular}
  \caption{Demographic information about the participants. M1-M18 tested MetaMask. W1-W18 tested iWallet (V1). W19-W26 tested iWallet (V2). 
  }~\label{demographic}
\end{table*}

\section{Pilot Study Results of iWallet (V1)}
\label{pilot}

The pilot results of iWallet (V1) showed that the educational materials helped users understand cryptocurrency and wallet concepts and terms. Most participants deemed the videos useful and were willing to watch them. However, some skipped videos due to their habit (e.g., W1), perceived uselessness (e.g., W10, \textit{``I skip everything but can still get it set up''}), and preferring text over videos (e.g., W15, \textit{``I'm a text person''}). 
W5 suggested making the videos mandatory, which could help novice users learn the important basics. W3 further suggested embedding a video to educate users on crypto-related laws. On average, our participants got 3.7 knowledge questions correctly before using the wallet, and that average increased to 4.4 after using the wallet, showing an 19\% improvement. 
In addition, nearly all blind users expressed that they preferred text instructions over video instructions since text were more accessible and saved time.

The security feature of confirming the receiving address was positively received by the participants. All of them expressed that the design made them more confident during the transaction process.
%

Our accessibility improvements led to less time for task completion for blind users. It took them 40 minutes on average to finish all the tasks,
while in the MetaMask evaluation, the blind users spent an average of 47.9 minutes on the tasks. 
However, we identified several accessibility challenges in our initial redesign. 
For example, our blind participants suggested putting explicit text explanations, such as ``wallet address,'' next to the elements, to help them more easily understand crypto concepts. The button labeling was not intuitive and informative enough for many blind users, e.g., when they clicked the Next button: \textit{``Next to what? More information is needed.''} 
Our blind participants further expressed that videos should be complemented by text summarizing their content, which was a favored medium for them to save time. 
The option of downloading the encrypted version of the secret recovery phrase was deemed helpful for blind users. However, we did not prioritize the option of uploading the secret recovery phrase in the account importing process. W11 expected the upload option on the left in the account importing page  but only saw the ``typing the phrase'' option. She failed to import her wallet since she typed in the encrypted phrase which was downloaded during the onboarding instead of the plain text phrase.
As a result of this inconsistent design, six out of eight blind participants were unable to import their wallets successfully. This highlighted the importance of consistency of function layout, especially for screen reader users who had more difficulty navigating. The accessibility glitches explained our relatively low average score \rev{(65 out of 100)} in the SUS survey. 

\end{document}